\documentclass[twocolumn,english,twocolumn,english,aps,prl,floats,showpacs]{revtex4}
\usepackage[T1]{fontenc}
\usepackage[latin9]{inputenc}
\usepackage{bm}
\usepackage{amsmath}
\usepackage{amssymb}
\usepackage{graphicx}
\usepackage{esint}

\makeatletter
\@ifundefined{textcolor}{}
{%
 \definecolor{BLACK}{gray}{0}
 \definecolor{WHITE}{gray}{1}
 \definecolor{RED}{rgb}{1,0,0}
 \definecolor{GREEN}{rgb}{0,1,0}
 \definecolor{BLUE}{rgb}{0,0,1}
 \definecolor{CYAN}{cmyk}{1,0,0,0}
 \definecolor{MAGENTA}{cmyk}{0,1,0,0}
 \definecolor{YELLOW}{cmyk}{0,0,1,0}
}

\usepackage{babel}

\usepackage{babel}

\usepackage{babel}

\usepackage{babel}

\makeatother

\usepackage{babel}
\begin{document}

\title{Persistent current in a 2D Josephson junction array wrapped around
a cylinder}

\author{D. A. Garanin and E. M. Chudnovsky}

\affiliation{Physics Department, Lehman College, The City University of New York,
250 Bedford Park Boulevard West, Bronx, NY 10468-1589, U.S.A.}

\date{\today}
\begin{abstract}
We study persistent currents in a Josephson junction array wrapped
around a cylinder. The $T=0$ quantum statistical mechanics of the
array is equivalent to the statistical mechanics of a classical $xy$
spin system in 2+1 dimensions at the effective temperature $T^{*}=\sqrt{2JU}$,
with $J$ being the Josephson energy of the junction and $U$ being
the charging energy of the superconducting island. It is investigated analytically and numerically
on lattices containing over one million sites. For weak disorder and  $T^{*}\ll J$
the dependence of the persistent current on disorder and $T^{*}$ computed numerically
agrees quantitatively with the analytical result derived within the spin-wave
approximation. The high-$T^{*}$ and/or strong-disorder behavior is dominated by instantons
corresponding to the vortex loops in 2+1 dimensions. The current becomes
destroyed completely at the quantum phase transition into the Cooper-pair
insulating phase.
\end{abstract}


\maketitle

\section{Introduction}

Persistent currents in closed chains of Josephson junctions (JJ) have
been studied for more than two decades (see, e.g. Refs.\ \onlinecite{Choi-1993,Matveev,Pop,Rastelli,garchu-PRB16}
and references therein). Being conceptually similar to small superconducting
rings they provide a testing ground for models of quantum phase slips
\cite{Zaikin,Golubev,Oreg,Semenov-PRB13} and superconductor-insulator
transition (SIT) \cite{Bradley,Korshunov,Chow}. This research has
significantly intensified in recent years due to advances in manufacturing
of nanostructures \cite{Mooij-2015} and the renewed interest to quantum
phase transitions \cite{Girvin,Sachdev} inspired in part by the prospects
of applications of quantum circuitry \cite{Ioffe,Gladchenko,Manucharyan}.

\begin{figure}[htbp!]
\includegraphics[width=10cm]{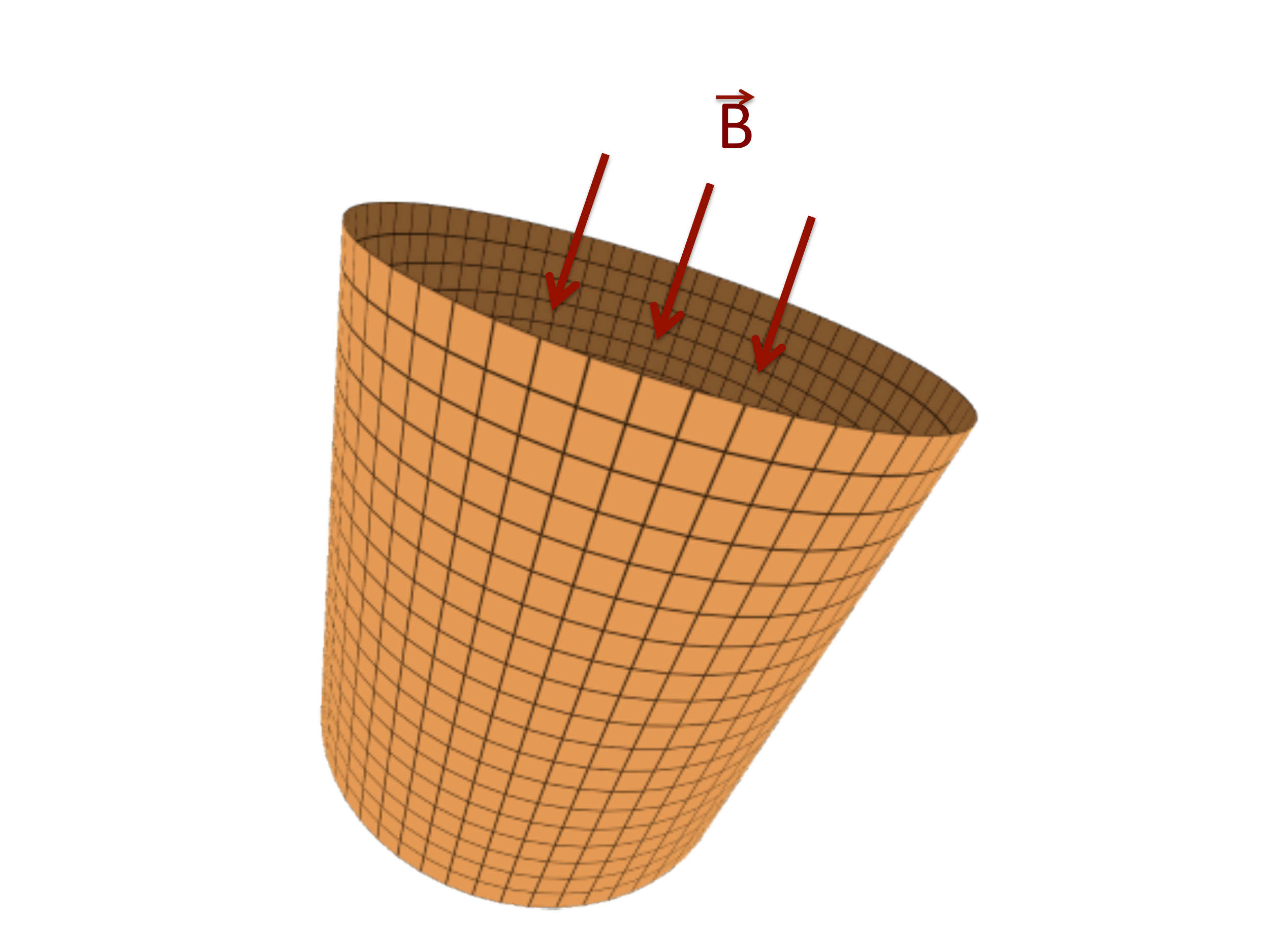} \caption{Color online: A $2D$ JJ array wrapped around a cylinder that encloses
the magnetic flux.}

\label{Fig:JJcylinder}
\end{figure}

%\begin{center}
There exist even a greater volume of work on SIT in disordered ultrathin
films (see, e.g. Ref.\ \onlinecite{Goldman} and references therein).
Similar to superconducting rings, various mechanisms of SIT have been
modeled by two-dimensional JJ arrays \cite{Fazio-Schon,Imry,Vinokur,Altshuler,EC}.
To have a persistent current in a $2D$ JJ array the latter should
be closed into a cylindrical surface that encloses the magnetic flux,
see Fig.\ \ref{Fig:JJcylinder}. In this Letter we argue that studies
of persistent currents in such a system provides another avenue for
testing the theory of quantum phase transitions. It may also be relevant
to properties of an ultrathin superconducting film deposited on a
cylinder and properties of a superconducting topological insulator \cite{Bernevig}.

%\par\end{center}

%\begin{center}
A system that bears some relevance to the JJ array wrapped around
a cylinder is a JJ ladder made of capacitively coupled one-dimensional
JJ rings, e.g., JJ necklaces stuck together. It was studied theoretically
by the numerical density-matrix renormalization group with an emphasis
on the role of excitons \cite{Lee-PRB03}. In this Letter we are taking
a different approach. When the dynamics of the $2D$ JJ array is dominated
by capacitances of superconducting islands its $T=0$ quantum statistical
mechanics is equivalent \cite{Bradley,Choi-1993,Wallin-94,Girvin,Kim-08}
to the statistical mechanics of the classical $xy$ spin system in
$2+1$ dimensions at the effective temperature $T^{*}=\sqrt{2JU}$,
with $J$ being the Josephson energy of the junction and $U$ being
the charging energy of the superconducting island. The advantage of
such a mapping is its suitability for large-scale Monte Carlo (MC)
studies.
%\par\end{center}

To make this problem relevant to experimental systems it must also
include disorder. Positional disorder in a planar JJ array is known
to give rise to random phases when the array is placed in the transverse
magnetic field \cite{Granato}. The resulting ``gauge'' or ``Bose''
glass has been intensively studied by analytical \cite{Rubinstein,KorNat,Li,Gupta}
and numerical \cite{Tang,Kim-06,Kim-08} methods. Phase disorder has
been found to have a significant effect on the SIT. The model with
the transverse magnetic field would not apply to the JJ array wrapped
around a cylinder. However, static random phases in Josephson junctions
can also be generated by other mechanisms, e.g., by the broken time-reversal
symmetry in the presence of the magnetic moments \cite{Buzdin-PRL08,Buzdin-PRL09}.

At first we neglect disorder. Let $\theta_{ij}$ be the phase of the
superconducting order parameter $\Psi=|\Psi|\exp(i\theta)$ at the
$ij$-th superconducting island, with $i=1,2,...,N$ denoting the
islands in the $j$-th ring ($j=1,2,...,N$) of the $2D$ JJ array
shown in Fig.\ \ref{Fig:JJcylinder}. The Josephson energy of the
array is \cite{Tinkham}
\begin{eqnarray}
 &  & E_{J}=J\sum_{ij}\left[1-\cos\left(\theta_{i,j+1}-\theta_{ij}\right)\right]+\nonumber \\
 &  & J\sum_{ij}\left[1-\cos\left(\theta_{i+1,j}-\theta_{ij}+\frac{2\pi}{\Phi_{0}}\int_{i}^{i+1}{\bf A}\cdot d{\bf l}_{j}\right)\right],\nonumber \\
\label{H-2dRing}
\end{eqnarray}
where the vector potential ${\bf A}$ is due to the magnetic flux
$\Phi$ enclosed by the cylinder.

Summation along each $j$-th ring gives
\begin{equation}
\sum_{i}\left(\theta_{i+1,j}-\theta_{ij}\right)=2\pi m_{j},\quad\frac{2\pi}{\Phi_{0}}\sum_{i}\int_{i}^{i+1}{\bf A}\cdot d{\bf l}_{j}=2\pi\phi,\label{sum}
\end{equation}
where $\phi\equiv\Phi/\Phi_{0}$, $\Phi_{0}=h/(2e)$ is the flux quantum,
and $m_{j}$ is an integer. This allows one to write
\begin{eqnarray}
 &  & E_{J}=J\sum_{ij}\left[1-\cos\left(\tilde{\theta}_{i,j+1}-\tilde{\theta}_{ij}\right)\right]+\nonumber \\
 &  & J\sum_{ij}\left[1-\cos\left(\tilde{\theta}_{i+1,j}-\tilde{\theta}_{ij}+\frac{2\pi\left(\phi+m_{j}\right)}{N}\right)\right],\label{eq:EJ_def}
\end{eqnarray}
where the reduced phases $\tilde{\theta}$ are defined in such a way
that the change of $\tilde{\theta}_{i}$ around the $j$-th ring is
zero. The total accumulation of the original phase $\theta_{ij}$
in a closed path around the cylinder is accounted for by the quantum
number $m_{j}$. The persistent current in the cylinder is given by
\begin{equation}
I=\frac{d\left\langle E_{J}\right\rangle }{d\Phi}=\frac{1}{\Phi_{0}}\frac{d\left\langle E_{J}\right\rangle }{d\phi},\label{I-GS}
\end{equation}

\begin{center}
\begin{figure}[htbp!]
\includegraphics[width=8cm]{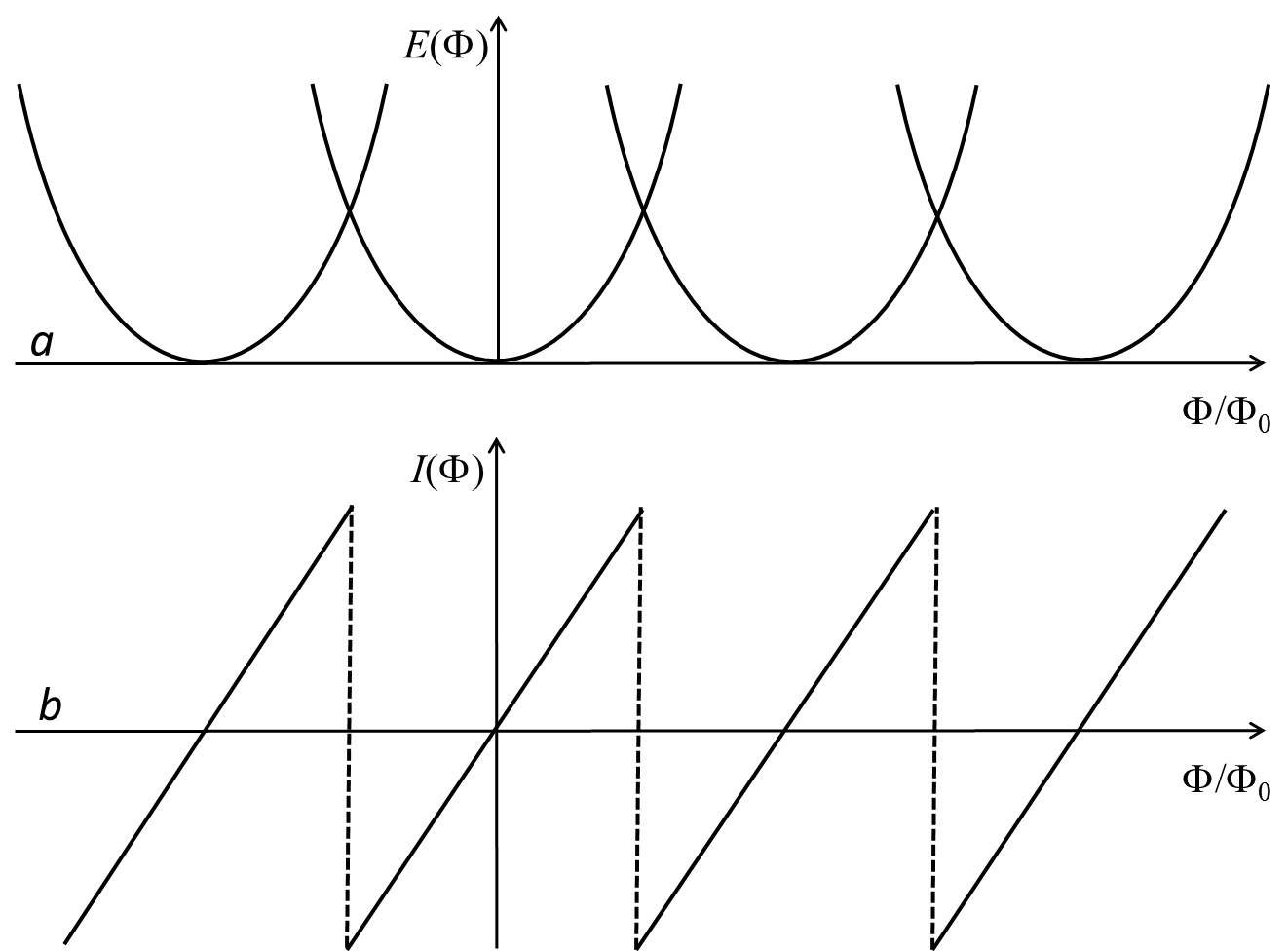} \caption{($a$) $m$-branches of the ground-state energy $E_{J}^{(0)}$. ($b$)
$m$-branches of the persistent current. }

\label{Fig:m-branches}
\end{figure}

\par\end{center}

The energy minimum corresponds to all $\tilde{\theta}_{ij}$ and all
$m_{j}$ being the same ($m_{j}=m$), leading to the ground state
\begin{equation}
E_{J}^{(0)}=JN^{2}\left[1-\cos\left(\frac{2\pi\left(\phi+m\right)}{N}\right)\right]\cong2\pi^{2}J(\phi+m)^{2},\label{GS}
\end{equation}
the last expression being the large-$N$ case. Branches of $E_{J}^{(0)}(\Phi)$
and $I^{(0)}(\Phi)={dE_{J}^{(0)}}/{d\Phi}$ for different values of
$m$ are shown in Fig.\ \ref{Fig:m-branches}$a$ and Fig.\ \ref{Fig:m-branches}$b$
respectively. When $\phi=n+1/2$, with $n$ being an integer, that
is for $\Phi=(n+1/2)\Phi_{0}$, the ground state is degenerate, $E_{J}(n,m)=E_{J}(n,m'=-2n-m-1)$.
For, e.g., half a fluxon, $\Phi=\Phi_{0}/2$, the current has the
same absolute value but flows in opposite directions for $m=0$ and
$m=-1$.

The dynamics of the JJ array is due to the electrical charging of
the superconducting islands by the excess (or lack) of Cooper pairs
$n_{ij}$ at the $ij$-th site. It is determined by the finite capacitances
of the islands to the ground and the capacitances of the junctions.
Different limits, with both capacitances present, can be achieved
in experiment and have been studied in literature, see, e.g., Ref.
\onlinecite{Rastelli} and references therein. In this paper we
are considering the limit in which the capacitances of the islands,
$C$, greatly exceed the capacitances of the junctions. In this case
the charging energy is given by
\begin{equation}
E_{C}=\sum_{ij}Un_{ij}^{2}=\frac{\hbar^{2}}{4U}\sum_{ij}\left(\frac{d\theta_{ij}}{dt}\right)^{2}\label{EC}
\end{equation}
where $U=(2e)^{2}/(2C)$. The second of Eq.\ (\ref{EC}), which plays
the role of the kinetic energy, is obtained by noticing that $n_{ij}$
and $\theta_{ij}$ are canonically conjugated variables, $n_{ij}=-i{d}/{d\theta_{ij}}$,
so that $i\hbar{d\theta_{ij}}/{dt}=[\theta_{i},E_{C}]=2iUn_{i}$.
Finite $U$ permits quantum tunneling between the current states that
correspond to $E_{J}(n,m)$ and $E_{J}(n,m')$.

The Lagrangian of the model is ${\cal {L}}=E_{C}-E_{J}$. Its quantum
mechanics is formulated in terms of the path integral
\begin{equation}
{\cal {I}}=\prod_{i}\int D\{\theta_{i}(\tau)\}e^{-S_{E}/\hbar}
\end{equation}
where $\tau=it$ and $S_{E}=\int d\tau{\cal L}$ is the Euclidean
action with
\begin{eqnarray}
{\cal L} & = & \frac{\hbar^{2}}{4U}\sum_{ij}\left(\frac{d\theta_{ij}}{d\tau}\right)^{2}+J\sum_{ij}\left[1-\cos\left(\theta_{i,j+1}-\theta_{ij}\right)\right]\nonumber \\
 & + & J\sum_{ij}\left[1-\cos\left(\theta_{i+1,j}-\theta_{ij}+\frac{2\pi\phi}{N}\right)\right].
\end{eqnarray}
Statistical mechanics of this quantum model at $T=0$ is equivalent
\cite{Bradley,Choi-1993,Wallin-94,Girvin,Kim-08} to the statistical
mechanics of the classical model in 2+1 dimensions at a non-zero temperature
$T^{*}=\sqrt{2JU}$, described by the partition function
\begin{equation}
Z=\prod_{i}\int D\{\theta_{i}(\tau)\}e^{-\mathcal{H}_{2+1}/T^{*}}\label{eq:Z_def}
\end{equation}
with $\mathcal{H}_{1+1}=-\frac{1}{2}\sum_{\mathbf{rr}'}J_{\mathbf{rr}'}\cos\left(\theta_{\mathbf{r}'}-\theta_{\mathbf{r}}+\phi_{\mathbf{rr}'}\right)$,
where $\mathbf{r}$ is a discrete three-dimensional vector $\mathbf{r}=(i,j,l)$
representing the space-time lattice, while $J_{\mathbf{rr}'}=J$ for
the nearest neighbors and zero otherwise. In the numerical work we
use the $N\times N\times N$ lattice with the $l$ direction corresponding
to the imaginary time and periodic boundary consitions. Non-zero phase
shifts are given by $\phi_{i,j,l;i\pm1,j,l}=\pm2\pi\phi/N$. Notice
that, in principle, the periodic boundary condition imposed on the
imaginary time introduces a finite physical temperature into the original
quantum problem, $T\sim T^{*}/N$. At large $N$ the effect of that
temperature on the persistent current can be ignored.

The statistical model presented above can be reformulated in terms
of the three-component classical spin vectors of the $3D$ $xy$ model
at temperature $T^{*}=\sqrt{2JU}$ that describes the strength of
quantum fluctuations. That model has a ferromagnetic-paramagnetic
phase transition at $T^{*}=T_{c}$ that in the original model corresponds
to the quantum phase transition into the Cooper-pair insulator state
in which the islands connected by Josephson junctions maintain their
superconductivity but no Josephson current can circulate around the
cylinder. The natural way to test this prediction is to study the
dependence of the persistent current on $U$.

Phase slips corresponding to quantum tunneling between different $m$
require formation of vortex loops in 2+1 dimensions. At small $U$
satisfying $T^{*}=\sqrt{2JU}\ll T_{c}\sim J$ such loops nucleate
with exponentially small probability. Consequently, at a small $T^{*}$,
if one induces a persistent current by placing the cylinder with the
JJ array in the magnetic field, the phase slips may not occur on the
time scale of the experiment. In this case all $m_{j}$ in Eq.\ \ref{eq:EJ_def}
are the same, $m_{j}=m$, and the persistent current computed with
the help of Eq. (\ref{I-GS}) and the symmetry becomes
\begin{equation}
I=\frac{2\pi NJ}{\Phi_{0}}\sin\left[\frac{2\pi\left(\phi+m\right)}{N}\right]\langle\cos(\tilde{\theta}_{i,j,l}-\tilde{\theta}_{i+1,j,l})\rangle.\label{I-av}
\end{equation}
We now recall that the statistical mechanics of our model is that
of the $3D$ $xy$ model at $T=T^{*}$, for which the low-temperature
(spin-wave) result for the cubic lattice is $\left\langle \cos\left(\theta_{\mathbf{r}}-\theta_{\mathbf{r}+\boldsymbol{\delta}}\right)\right\rangle =1-T^{*}/(6J)$,
with $\boldsymbol{\delta}$ being the nearest neighbor in any direction.
This gives for large $N$
\begin{equation}
I\cong\frac{\left(2\pi\right)^{2}J}{\Phi_{0}}(\phi+m)\left(1-\frac{T^{*}}{6J}\right).\label{eq:I_SWT}
\end{equation}

We shall now consider the effect of quenched disorder by adding a
static random phase $\phi_{ij}$ to the phase difference, $\theta_{i,j+1}-\theta_{ij}$,
between neighboring superconducting islands $i$ and $j$ in Eq.\ (\ref{H-2dRing}).
In the models of planar JJ arrays such random phase can be generated
by the positional disorder in the $2D$ lattice of Josephson junctions
in the presence of the transverse magnetic field \cite{Granato,KorNat,Li,Gupta,Tang,Kim-06,Kim-08}.
In our case of a JJ array wrapped around a cylinder, random phases
can be generated due to, e.g., anomalous Josephson effect in the presence
of magnetic moments \cite{Buzdin-PRL08,Buzdin-PRL09}.

As was shown in Ref.\ \onlinecite{Rubinstein} the continuous counterpart
of the model with quenched randomness corresponds to the interaction
of the continuous phase order parameter $\theta({\bf r})$ with a
static random field ${\bf q}({\bf r})=\phi_{ij}\hat{r}_{ij}/a$ ($a$
being the lattice spacing), described by
\begin{equation}
E_{J}=\frac{1}{2}J\int d^{2}r[{\bm{\nabla}}\theta({\bf r})-{\bf q}({\bf r})]^{2}.\label{eq:EJ_cont}
\end{equation}
The Imry-Ma argument \cite{I-M} favors the destruction of the long-range
order in less than four dimensions by a weak static random field interacting
with the order parameter directly, since at $d<4$ such interaction,
regardless of strength, dominates the energy at large distances. Crucial
to that argument, however, is the formation of topological defects
which makes the order more robust \cite{PGC-PRL2014}. In our case,
random field in Eq.\ (\ref{eq:EJ_cont}) interacts with the gradient
of the order parameter, which further diminishes its effect at large
distances. One should expect, therefore, that the initially ordered
state will not be destroyed by weak quenched randomness at low $T^{*}$.

The persisent current in Eq.\ (\ref{I-av}) should now be calculated
with account of averaging over phase fluctuations generated by the
random field. Introducing random phases $\phi_{ij}$ into Eq.\ (\ref{eq:EJ_def}),
similarly to the above in the limit of large $N$ we obtain
\begin{equation}
I=\frac{(2\pi)^{2}J\left(\phi+m\right)}{\Phi_{0}}\left\langle \cos\left[\tilde{\theta}_{i+1,j,l}-\tilde{\theta}_{i,j,l}+\phi_{ij}\right]\right\rangle .
\end{equation}
In the case $|\phi_{ij}|\ll1$ and $|\tilde{\theta}_{i+1,j,l}-\tilde{\theta}_{i,j,l}|\ll1$
the continuous model of Eq. (\ref{eq:EJ_cont}) yields
\begin{equation}
I=\frac{(2\pi)^{2}J\left(\phi+m\right)}{\Phi_{0}}\left\{ 1-\frac{1}{2}a^{2}\left\langle [{\bm{\nabla}}\theta({\bf r})-{\bf q}({\bf r})]^{2}\right\rangle \right\} .
\end{equation}
The effects of low temperature and weak static randomness must be
additive. It suffices, therefore, to compute the contribution of quenched
randomness at $T^{*}=0$ just minimizing the energy, Eq. (\ref{eq:EJ_cont}).
The extrema of the latter satisfy the equation
\begin{equation}
{\bm{\nabla}}^{2}\theta({\bf r})={\bm{\nabla}}\cdot{\bf q}({\bf r})
\end{equation}
having the solution
\begin{equation}
\theta({\bf r})=\int d^{2}r'G({\bf r}-{\bf r}'){\bm{\nabla}}\cdot{\bf q}({\bf r}')=\int d^{2}r'{\bf q}({\bf r}')\cdot{\bm{\nabla}}G({\bf r}-{\bf r}'),
\end{equation}
where $G({\bf r})=(2\pi)^{-1}\ln(r/a)$ is the Green function of the
$2D$ Laplace equation, ${\bm{\nabla}}^{2}G({\bf r})=\delta({\bf r})$.
With the help of this equation one obtains
\begin{eqnarray}
 &  & {\nabla}_{\alpha}\theta({\bf r})-q_{\alpha}({\bf r})=\nonumber \\
 &  & \int d^{2}r'q_{\beta}({\bf r}')\left[{\nabla}_{\alpha}\nabla_{\beta}G({\bf r}-{\bf r}')-\delta_{\alpha\beta}\delta({\bf r}-{\bf r}')\right],
\end{eqnarray}
so that
\begin{eqnarray}
 &  & [{\bm{\nabla}}\theta({\bf r})-{\bf q}({\bf r})]^{2}=\int d^{2}r'\int d^{2}r''q_{\beta}({\bf r}')q_{\gamma}({\bf r}'')\nonumber \\
 &  & \times\left[{\nabla}_{\alpha}\nabla_{\beta}G({\bf r}-{\bf r}')-\delta_{\alpha\beta}\delta({\bf r}-{\bf r}')\right]\nonumber \\
 &  & \times\left[{\nabla}_{\alpha}\nabla_{\gamma}G({\bf r}-{\bf r}'')-\delta_{\alpha\gamma}\delta({\bf r}-{\bf r}'')\right].
\end{eqnarray}

We shall assume that
\begin{equation}
\langle q_{\beta}({\bf r}')q_{\gamma}({\bf r}'')\rangle=\frac{1}{2}a^{2}q_{R}^{2}\delta_{\beta\gamma}\delta({\bf r}'-{\bf r}'').
\end{equation}
Then
\begin{eqnarray}
 &  & \left\langle [{\bm{\nabla}}\theta({\bf r})-{\bf q}({\bf r})]^{2}\right\rangle =\nonumber \\
 &  & \frac{1}{2}a^{2}q_{R}^{2}\int d^{2}r\left[{\nabla}_{\alpha}\nabla_{\beta}G({\bf r})-\delta_{\alpha\beta}\delta({\bf r})\right]^{2}\nonumber \\
 &  & =\frac{1}{2}a^{2}q_{R}^{2}\int\frac{d^{2}k}{(2\pi)^{2}}\left[k_{\alpha}k_{\beta}G({\bf k})+\delta_{\alpha\beta}\delta({\bf k})\right]^{2}\nonumber \\
 &  & =\frac{1}{2}a^{2}q_{R}^{2}\int\frac{d^{2}k}{(2\pi)^{2}}\left[\delta_{\alpha\beta}-\frac{k_{\alpha}k_{\beta}}{k^{2}}\right]^{2}\nonumber \\
 &  & =\frac{1}{2}a^{2}q_{R}^{2}\int\frac{d^{2}k}{(2\pi)^{2}}=\frac{1}{2}q_{R}^{2},
\end{eqnarray}
where we have used Fourier transforms: $G({\bf k})=-1/k^{2}$ and
$\delta({\bf k})=1$. Thus
\begin{equation}
I=\frac{(2\pi)^{2}J\left(\phi+m\right)}{\Phi_{0}}\left(1-\frac{1}{4}a^{2}q_{R}^{2}\right).
\end{equation}

The last step is to establish the relation between $q_{R}$ and $\phi_{R}$
used in the numerical experiment. If $\phi_{ij}$ are chosen randomly
between $-\phi_{R}$ and $\phi_{R}$, one has
\begin{equation}
a^{2}q_{R}^{2}=\frac{1}{2\phi_{R}}\int_{-\phi_{R}}^{+\phi_{R}}d\phi\,\phi^{2}=\frac{\phi_{R}^{2}}{3}.
\end{equation}
Inserting this into the above formula and combining it with Eq. (\ref{eq:I_SWT}),
one finally obtains the persistent current at low $T^{*}$ in the
presence of weak randomness
\begin{equation}
I=\frac{(2\pi)^{2}J\left(\phi+m\right)}{\Phi_{0}}\left(1-\frac{T^{*}}{6J}-\frac{\phi_{R}^{2}}{12}\right).\label{eq:I_anal_final}
\end{equation}
As we shall see, this formula agrees well with numerical results.

\begin{figure}[htbp!]
\includegraphics[width=9cm]{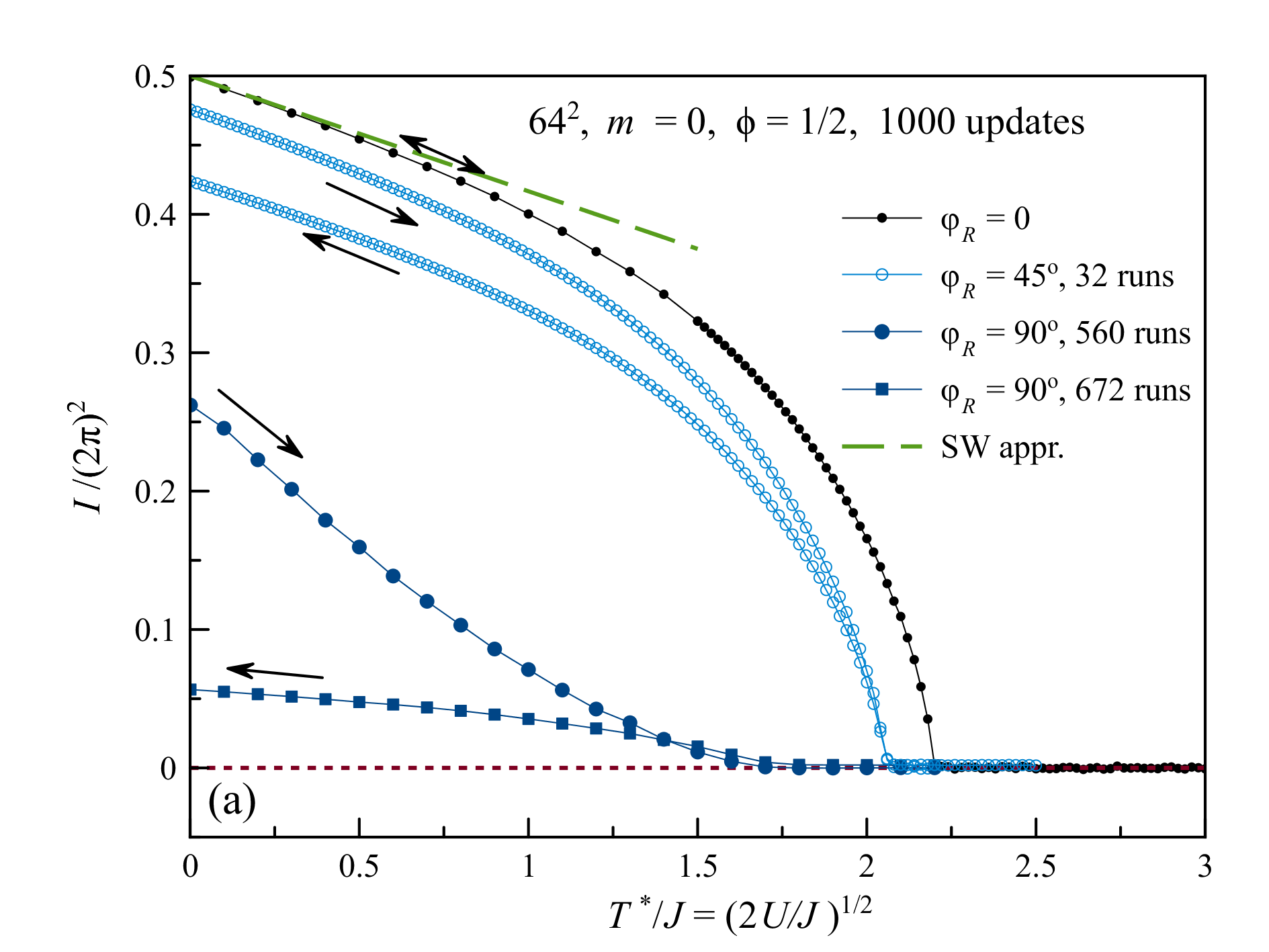}

\includegraphics[width=9cm]{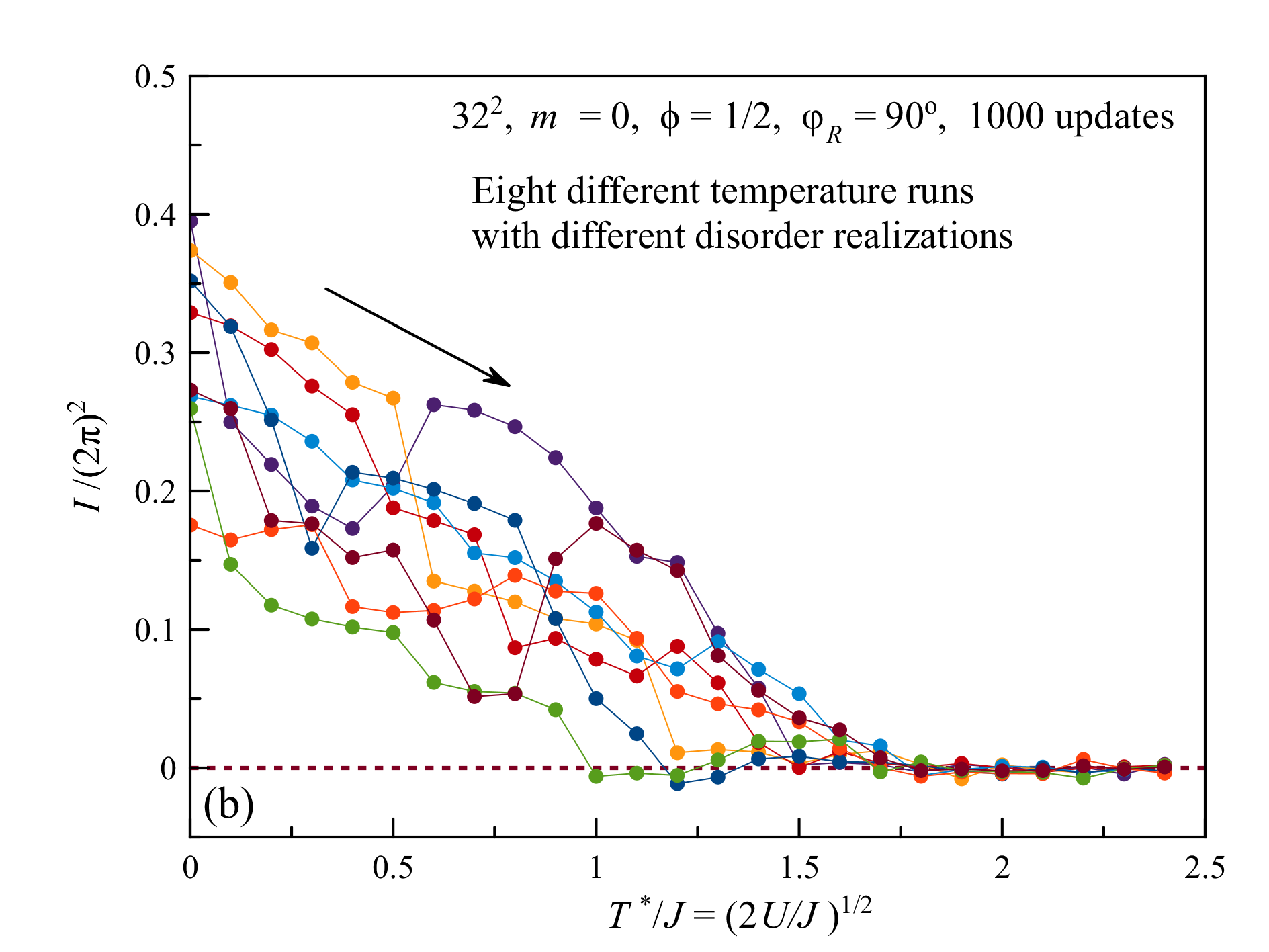}

\caption{Color online: Effective-temperature dependence of the persistent current.
(a) Averaged over disorder realizations (runs). (b) Individual runs
showing strong fluctuations for strong disorder. }

\label{Fig_3}
\end{figure}

Higher $T^{*}$ and $\phi_{R}$ tend to create topological defects.
Destruction of the persistent current by the latter and spin waves
can be investigated numerically using the equivalent magnetic model
of two-component vectors $\mathbf{s}$ of fixed length instead of
the grain phase $\theta$.

The case $T^{*}=0$ is classical, so that one can minimize the energy
of the $2D$ system of Josephson grains by the method of Ref.\ \onlinecite{garchu-PRB16}
that uses successive rotations of the grain's vectors $\mathbf{s}$
into the direction of the effective field $\mathbf{H}_{\mathrm{eff}}$
with probability $\alpha$ and overrelaxation (reflecting $\mathbf{s}$
with respect to $\mathbf{H}_{\mathrm{eff}}$) with probability $1-\alpha$.
Small $\alpha$ provide the highest efficiency. In this work we used
$\alpha=0.01$ everywhere. The results of energy minimization are
averaged over realizations of the disorder $\varphi_{ij}$. For this,
many runs were done until averages stabilized and smooth curves, such
as $I(T^{*})$, were obtained.

For non-zero $T^{*}$ the original $2D$ quantum problem was solved
as the effective classical $3D$ problem with the effective temperature
$T^{*}$ using the standard Monte Carlo procedure. For each classical
spin (Josephson grain) we used the Monte Carlo update with probability
$\alpha=0.01$ and overrelaxation with probability $1-\alpha$. For
each parameter value ($T^{*}$ or $\varphi_{R}$) at least 1000 system
updates were performed. This was sufficient for local equilibration and
averaging over thermal fluctuations. On top of it, averaging over
realizations of disorder (runs) was performed.

In the case of strong disorder the system demonstrates glassy properties,
so that relaxation leads to different final states depending on the
initial state and other factors. Here Monte Carlo routine does not
lead to global equilibration for $T^{*}\lesssim J$. To average out
glassy fluctuations, one has to perform many runs with different realizations
of disorder. The numerical problem for the persistent current $I$
is tougher than computation of the magnetization of a ferromagnet.
With a given number of fluxons $\phi$, the current is inversely proportional
to the length of the rings $N$ and directly proportional to the number
of the rings $N$, so that $I$ is independent of $N$, Eq. (\ref{eq:I_anal_final}).
Thus increasing the system size does not lead to strong suppression
of fluctuations, as in the case of the magnetization. Performing a
large number of repeated measurements (runs) is the only way to beat
fluctuations.

Fig. \ref{Fig_3}a shows dependence of the persistent current $I$
on the effective temperature $T^{*}$, obtained by increasing or decreasing
$T^{*}$ in small steps starting from the collinear spin state (same
phase everywhere, $m=0$). In the absence of disorder the process
of temperature change is reversible and Eq. (\ref{eq:I_SWT}) is a
good approximation at $T^{*}/J\ll1$. The current $I$ vanishes at
$T^{*}/J=2.22$ that corresponds to the ferromagnetic transition in
$3D$ $xy$ model. In contrast to the magnetization, no finite-size
effects are seen, the curves being practically the same for $N=32$,
64, 128. For $\phi_{R}=45^{\circ}$ there is a hysteresis and the
transition point is moving down, $T^{*}/J=2.06$. For $\phi_{R}=90^{\circ}$
hysteresis is very strong and the transition point is difficult to detect. Fig. \ref{Fig_3}b
reflects strong fluctuations from run to run for such strong disorder.

Fig. \ref{Fig_4} shows the dependence of the persistent current on the disorder strength $\varphi_{R}$
at $T^{*}=0$, computed for the $2D$ classical model, and for $T^{*}/J=1$.
All results are obtained starting from the collinear initial condition
(CIC) for each $\varphi_{R}$ value. The two curves cross
because of glassy effects and different computational methods used. At
$T^{*}=0$ minimization of the energy of the $2D$ system occurs faster
and better than Monte Carlo in $3D$, leading to slightly lower energies.
Perfect match of the curves for different sizes shows the absence of size
effects. Eq. (\ref{eq:I_anal_final}) works well for small disorder,
actually up to $\varphi_{R}\simeq60^{\circ}$. In this region vortex
loops begin to pop up and vorticity $f_{V}$ (see, e.g., Ref.\ \cite{garchu-PRB16})
starts to grow from zero. At $\varphi_{R}\simeq100^{\circ}$ the persistent
current practically vanishes.

\begin{center}
\begin{figure}[htbp!]
\includegraphics[width=9cm]{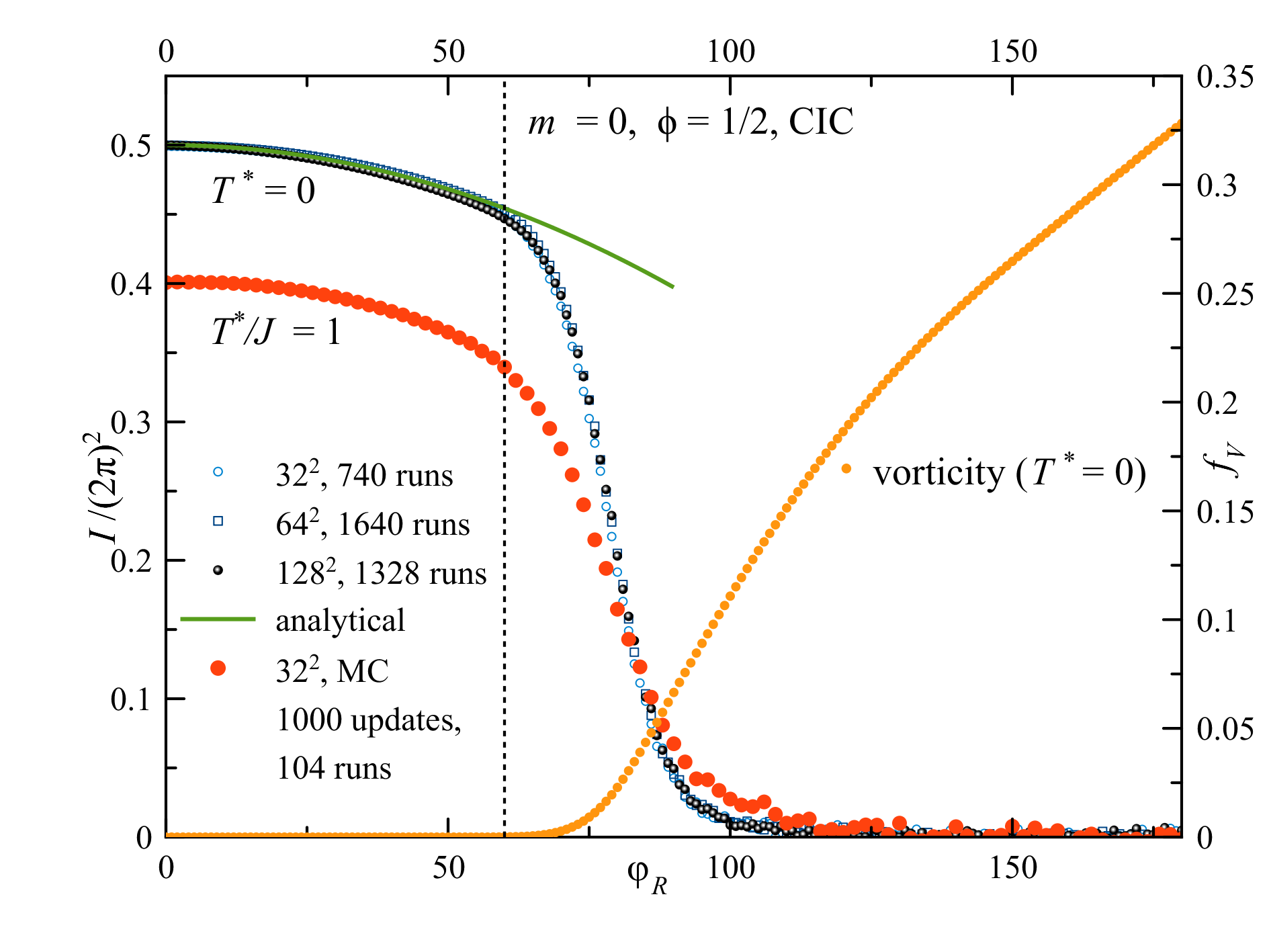} \caption{Color online: Disorder dependence of the persistent current and vorticity
at different effective temperatures. }
\label{Fig_4}
\end{figure}
\par\end{center}

In Conclusion, we have proposed a novel system for the study of quantum phase transitions: Josephson junction array wrapped around a cylinder. The dependence of the persistent current on the Josephson and charging energies and on the strength of disorder, computed analytically in the spin wave approximation, agrees well with numerical results. Cases of strong quantum fluctuations and strong disorder have been studied numerically on lattices having over one million sites. Quantum phase transition in the Josephson junction array wrapped around a cylinder is dominated by instantons corresponding to the vortex loops in 2+1 dimensions. Experimental study of such a system would be of great interest.

This work has been supported by the grant No. DE-FG02-93ER45487 funded
by the U.S. Department of Energy, Office of Science.

\end{document}